\documentclass[runningheads]{llncs}
\usepackage[margin=1.25in]{geometry}
\usepackage{cite}
\usepackage{stfloats}
\usepackage{amsmath,amssymb,amsfonts}
\usepackage{algorithm,algorithmicx}
\usepackage[noend]{algpseudocode}
\usepackage{graphicx}
\usepackage{textcomp}
\usepackage{hyperref}
\usepackage{multirow}
\usepackage[normalem]{ulem}

\DeclareMathOperator*{\argmin}{arg\,min}
\usepackage{tikz}
\usepackage{mathtools}
\usepackage{graphicx}
\newcommand{\norm}[1]{\left\lVert#1\right\rVert}

\usepackage{tikz}
\usepackage{mathtools}
\usetikzlibrary{shapes,arrows}
\usetikzlibrary{positioning}
\usetikzlibrary{fit}
\tikzstyle{decision} = [diamond, draw, fill=blue!20, 
    text width=4.5em, text badly centered, node distance=3cm, inner sep=0pt]
\tikzstyle{block} = [rectangle, draw, fill=gray!20, 
    text width=3em, text centered, rounded corners, minimum height=4em]
    \tikzstyle{dblock} = [rectangle, draw, dashed, fill=blue!20, 
    text width=3em, text centered, rounded corners, minimum height=4em]
    \tikzstyle{rblock} = [rectangle, draw, fill=gray!20, 
    text width=3em, text centered, rounded corners, minimum height=4em]
\tikzstyle{line} = [draw, -latex']
\tikzstyle{cloud} = [draw, ellipse,fill=red!20, node distance=3cm,
    minimum height=2em]
\tikzstyle{blank} = [node distance=3cm,
    minimum height=2em]  
\begin{document}

\title{Regularization-Agnostic Compressed Sensing MRI Reconstruction with Hypernetworks}
\titlerunning{Regularization-Agnostic CS-MRI Reconstruction}

\author{Alan Q. Wang\inst{1} \and
Adrian V. Dalca \inst{2, 3} \and
Mert R. Sabuncu\inst{1}} 

\authorrunning{A. Wang et al.}

\institute{School of Electrical and Computer Engineering, Cornell University \and Computer Science and Artificial Intelligence Lab, Massachusetts Institute of Technology \and A.A. Martinos Center for Biomedical Imaging, Massachusetts General Hospital, Harvard Medical School.}

\maketitle              
\begin{abstract}
Reconstructing under-sampled $k$-space measurements in Compressed Sensing MRI (CS-MRI) is classically solved with regularized least-squares. Recently, deep learning has been used to amortize this optimization by training reconstruction networks on a dataset of under-sampled measurements. 
Here, a crucial design choice is the regularization function(s) and corresponding weight(s). 
In this paper, we explore a novel strategy of using a hypernetwork to generate the parameters of a separate reconstruction network as a function of the regularization weight(s), resulting in a regularization-agnostic reconstruction model. 
At test time, for a given under-sampled image, our model can rapidly compute reconstructions with different amounts of regularization. We analyze the variability of these reconstructions, especially in situations when the overall quality is similar. Finally, we propose and empirically demonstrate an efficient and data-driven way of maximizing reconstruction performance given limited hypernetwork capacity. Our code is publicly available at \url{https://github.com/alanqrwang/hyperrecon}.
\keywords{Compressed Sensing MRI  \and Hypernetworks \and Unsupervised Reconstruction.}
\end{abstract}
\section{Introduction}
\label{sec:introduction}
Magnetic resonance imaging (MRI) can be accelerated by under-sampling $k$-space -- a technique known as Compressed Sensing MRI (CS-MRI)~\cite{lustig}. This yields a well-studied ill-posed inverse problem. Classically, this problem is reduced to iteratively minimizing a regularized regression cost function on each collected measurement set~\cite{fista,admmboyd,primaldualchambolle,combettes2009proximal,daubechies2003iterative,Ye2019}. In contrast, learning-based amortized optimization trains a neural network to minimize the cost function over a dataset of under-sampled measurements~\cite{shu2018amortized,wang2020amortized}. This method has been shown to significantly speed up computation time while outperforming traditional instance-based methods. 

However, both the classical and amortized methods suffer from a strong dependence on the regularization function, the weighting of which can vastly affect reconstructions. 
Thus, this hyperparameter must be carefully tuned to achieve optimal reconstruction quality. 
As a result, considerable time and resources can be spent setting the weight(s) of regularization, using methods such as grid-search, random search, or Bayesian optimization.

An emerging solution to the problem of hyperparameter tuning involves the use of hypernetworks which enable the user to jointly learn the model weights and optimal hyperparameters concurrently. Such methods have been applied broadly in learning models~\cite{brock2018smash,lorraine2018stochastic} and recently in medical image registration~\cite{hoopes2021hypermorph}. In this paper, we build on this line of thinking, applying it to the unsupervised reconstruction problem in CS-MRI and proposing a \textit{regularization-agnostic} strategy for performing amortized optimization. 

Specifically, our model uses a hypernetwork that takes as input a value (or values) for the regularization weight(s) and outputs the parameters of a reconstruction network. At test-time, arbitrary weight values can be provided and corresponding reconstructions will be efficiently computed. 
Thus, our method is capable of producing a range of reconstructions over the entire landscape of hyperparameter values, instead of a single reconstruction associated with a single hyperparameter value setting. We envision a workflow in which a user will select from many diverse reconstructions for further use based on visual inspection.

Additionally, we investigate different methods of sampling from the hyperparameter space during training. In particular, we examine two sampling procedures: uniform hyperparameter sampling (UHS) and a novel data-driven non-uniform hyperparameter sampling (DHS) strategy. We visualize and quantify the impact of sampling on the quality and range of reconstructions, using empirical results to demonstrate significant improvement when data-driven hyperparameter sampling is used during training.

\section{Background}

\subsection{Amortized Optimization of CS-MRI}

In the classical CS-MRI formulation, a reconstruction of the unobserved fully-sampled MRI scan\footnote{In this paper, we assume a single coil acquisition.} $\boldsymbol{x} \in \mathbb{C}^N$ is obtained by solving an ill-posed inverse problem via an optimization of the form
\begin{equation}
    \argmin_{\boldsymbol{x}} J(\boldsymbol{x},\boldsymbol{y}) + \sum_{i=1}^p \alpha_i \mathcal{R}_i(\boldsymbol{x})
    \label{eq:classical}
\end{equation}
for each under-sampled measurement $\boldsymbol{y} \in \mathbb{C}^M$, where $N$ is the number of pixels of the full-resolution grid and $M<N$ is the number of measurements. 

The first term, called the data-consistency loss, quantifies the agreement between the measurement vector $\boldsymbol{y}$ and reconstruction and can be defined as:
\begin{equation}
    J(\boldsymbol{x}, \boldsymbol{y}) = \norm{\mathcal{F}_u \boldsymbol{x} - \boldsymbol{y}}^2_2,
    \label{eq:dcloss}
\end{equation}
where $\mathcal{F}_u$ denotes the under-sampled Fourier operator.
The second term is composed of a sum of $p$ regularization functions, $\mathcal{R}_i(\boldsymbol{x})$, each weighted by a hyperparameter $\alpha_i \in \mathbb{R}_+$, that determines the degree of trade-off between the $p+1$ competing terms. The regularization term is often carefully engineered to restrict the solutions to the space of desirable images. Common choices include sparsity-inducing norms of wavelet coefficients~\cite{waveletnorm}, total variation (TV)~\cite{hdtv,1992PhyD60259R}, and their combinations~\cite{lustig,Ravishankar2020}.

A recently popular approach involves solving the instance-based optimization of Eq.~\eqref{eq:classical} with a neural network \cite{wang2020amortized}. For a neural network $G_\theta$ parameterized with $\theta$ and a training set $\mathcal{D}$ of under-sampled measurements $\boldsymbol{y}$, the problem reduces to:
\begin{equation}
    \arg\min_\theta \sum_{\boldsymbol{y} \in \mathcal{D}} \left[J(G_\theta(\boldsymbol{y}), \boldsymbol{y}) + \sum_{i=1}^p \alpha_i \mathcal{R}_i(G_\theta(\boldsymbol{y}))\right].
    \label{eq:amortized}
\end{equation}
This formulation can be viewed as an amortization of the instance-specific optimization of Eq.~\eqref{eq:classical}, via a neural network $G_\theta$~\cite{Balakrishnan2019,cremer2018inference,Gershman2014AmortizedII,marino2018iterative,shu2018amortized}. 

Amortized optimization provides several advantages over classical solutions. First, at test-time, it replaces an expensive iterative optimization procedure with a simple forward pass of a neural network.
Second, since the function $G_\theta$ is tasked with estimating the reconstruction for any viable input measurement vector $\boldsymbol{y}$ and not just a single instance, amortized optimization has been shown to act as a natural regularizer for the optimization problem~\cite{Balakrishnan2019,shu2018amortized}.

\subsection{Hypernetworks}
Originally introduced for achieving weight-sharing and model compression \cite{ha2016hypernetworks}, hypernetworks take as input a set of hyperparameters that are in turn converted to the weights of the network that solves the main task, such as classification.
In this framework, the parameters of the hypernetwork, and not the main-task network, are learned.
This idea has found numerous applications including neural architecture search \cite{brock2018smash,zhang2019graph}, Bayesian neural networks \cite{krueger2018bayesian,pmlr-v95-ukai18a}, multi-task learning \cite{pan2018hyperstnet,shen2017meta,Klocek2019}, and hyperparameter optimization \cite{lorraine2018stochastic}. 
In an application related to ours, for example, hypernetworks have been used as a means of replacing cross-validation for finding the optimal level of weight decay on MNIST classification~\cite{lorraine2018stochastic}. The authors propose a training scheme which allows the hypernetwork to output the optimal weights of a main network as a function of the weight decay parameter. Similarly, in parallel work hypernetworks were used to build a hyperparameter-invariant medical image registration framework~\cite{hoopes2021hypermorph}. 

\section{Proposed Method}
\begin{figure}[t]
\centering
\includegraphics[width=0.7\textwidth]{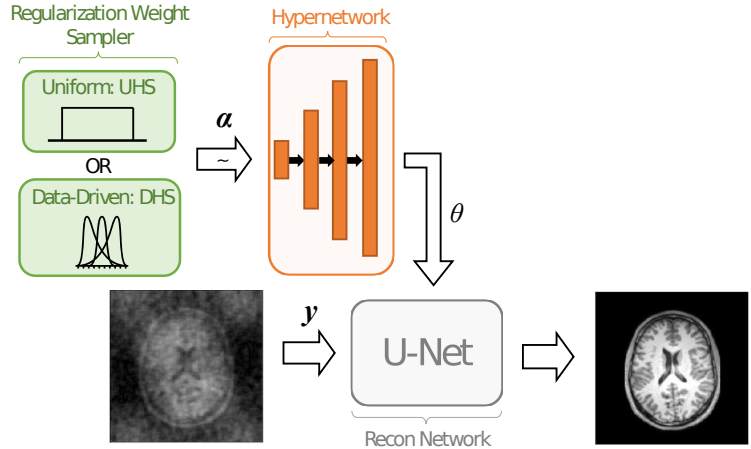}
\caption{Proposed model. Reconstruction network takes as input the under-sampled measurement $\boldsymbol{y}$ and outputs reconstruction $\boldsymbol{\hat{x}}$, while hypernetwork $H_\phi$ takes as input the hyperparameter $\boldsymbol{\alpha}$ and outputs weights $\theta$ for the reconstruction network. During training, $\boldsymbol{\alpha}$ is sampled from an either uniform (UHS) or data-driven (DHS) distribution.}
\label{fig:model-arch}
\end{figure}

Inspired by the prior applications of hypernetworks, we propose to use them for the CS-MRI reconstruction problem, where our goal is to produce a model that can efficiently compute a reconstruction that approximates the solution to Eq.~\eqref{eq:classical} for arbitrary regularization weight values.
We call this model \textit{regularization-agnostic}, and illustrate it in Fig.~\ref{fig:model-arch}.

\subsection{Regularization-Agnostic Reconstruction Network}
Suppose all regularization weights are gathered in a $p$-vector 
$$\boldsymbol{\alpha} = [\alpha_1, \alpha_2,..., \alpha_p]^T~\in~\mathbb{R}_+^p.$$ As before, let $G_\theta$ denote a main network which maps under-sampled measurements $\boldsymbol{y}$ to reconstructions $\boldsymbol{\hat{x}}$, whose parameters are denoted as $\theta \in \Theta$. We define a hypernetwork $H_\phi : \mathbb{R}_+^p \rightarrow \Theta$ that maps a regularization weight vector $\boldsymbol{\alpha}$ to the parameters $\theta$ of the main network $G_\theta$. During training we only learn $\phi$, the parameters of the hypernetwork.

Given a dataset $\mathcal{D}$ of under-sampled measurements~$\boldsymbol{y}$, the objective is to minimize the following loss with respect to $\phi$:
\begin{align}
     \arg\min_\phi \mathbb{E}_{\boldsymbol{\alpha} \sim p(\boldsymbol{\alpha})} \sum_{\boldsymbol{y} \in \mathcal{D}} &\left[J(G_{H_\phi(\boldsymbol{\alpha})}, \boldsymbol{y}) + 
    \sum_{i=1}^p \alpha_i \mathcal{R}_i\left(G_{H_\phi(\boldsymbol{\alpha})}(\boldsymbol{y})\right)\right].
    \label{eq:unsupervised}
\end{align}
%

In the hypothetical scenario of infinite hypernetwork capacity, the hypernetwork can capture a mapping of any input $\boldsymbol{\alpha}$ to the optimal $\theta^* = H_{\phi^*}(\boldsymbol{\alpha})$ that minimizes Eq.~\eqref{eq:amortized} with the corresponding $\boldsymbol{\alpha}$. The training of the hypernetwork is not sample limited, since one can draw as many hyperparameter samples as needed. Thus, overfitting for the hypernetwork is not a practical concern. However, in practice, we will have a limited capacity  hypernetwork and its learned parameters will in general depend on the assumed distribution for $\boldsymbol{\alpha}$. 
In a sense, while $G_\theta$ amortizes over the space of images, $H_\phi$ can be thought of as amortizing over the space of regularization weights. 

In general, the expression inside the braces of Eq.~\eqref{eq:unsupervised} can be manipulated such that the hyperparameter support is bounded to $\boldsymbol{\alpha} \in [0,1]^p$. For example, for one and two regularization weights, we can use:
\begin{align}
\label{eq:one-hyperparameter}
\begin{split}
    \mathcal{L}_{p=1}(\boldsymbol{y}, \boldsymbol{\alpha}) = (1-\alpha_1)J(\boldsymbol{\hat{x}}, \boldsymbol{y}) + 
    \alpha_1\mathcal{R}_1\left(\boldsymbol{\hat{x}}\right),
\end{split} \\
\label{eq:two-hyperparameter}
\begin{split}
    \mathcal{L}_{p=2}(\boldsymbol{y}, \boldsymbol{\alpha}) = \alpha_1 J(\boldsymbol{\hat{x}}, \boldsymbol{y}) + 
    (1-\alpha_1)\alpha_2 \ \mathcal{R}_1(\boldsymbol{\hat{x}}) + (1-\alpha_1)(1-\alpha_2)\ \mathcal{R}_2(\boldsymbol{\hat{x}}), 
\end{split}
\end{align}
where $\boldsymbol{\hat{x}} = G_{H_\phi(\boldsymbol{\alpha})}(\boldsymbol{y})$ denotes the model reconstruction. 

\subsection{Training}
\subsubsection{Uniform Hyperparameter Sampling (UHS):}
A straightforward strategy for training the hypernetwork involves sampling the regularization weights from a uniform distribution $p(\boldsymbol{\alpha}) = U[0,1]^p$ and an under-sampled measurement vector $\boldsymbol{y}$ for each forward pass during training. 
The gradients are then computed with respect to the loss evaluated at the sampled $\boldsymbol{\alpha}$ via a backward pass.
This corresponds to minimizing Eq.~\eqref{eq:unsupervised} with a uniform distribution for~$\boldsymbol{\alpha}$.
Algorithm~\ref{algorithm1} details the pseudo-code for the resulting training procedure.

However, the finite model capacity of the hypernetwork constrains the ability to achieve optimal loss for every hyperparameter value, particularly without resorting to training large and computationally expensive hypernetworks. 
In addition, sampling hyperparameters from the entire support $[0,1]^p$ might ``waste" model capacity on regularization weight values that have no hope of producing acceptable reconstructions, even if solved optimally. These two observations suggest that modifying the hyperparameter sampling distribution could lead to better performance by making better use of limited hypernetwork model capacity.

Unfortunately, in most real-world reconstruction scenarios, we don't have a good \textit{prior} distribution model for sampling the regularization weights. In the remainder of this section, we propose an unsupervised, data-driven sampling scheme which learns the optimal sampling region during training, and show in the Experiments section how this significantly improves reconstruction performance.

\begin{figure}[t]
\begin{minipage}[t]{0.49\textwidth}

\begin{algorithm}[H]
    \caption{\\UHS: Uniform Sampling}\label{algorithm1}
    Input: Dataset $\mathcal{D}$, $p$, $B$, $\gamma>0$ \\
    Output: Model weights $\phi$
    
    \begin{algorithmic}[1]
    \Repeat
        \State Sample $\{y_1, y_2,...,y_B\} \sim \mathcal{D}$
            \State $\alpha \leftarrow \alpha^i \sim U[0,1]^p \ \text{for} \ i=1,...,B$
            \State $\phi \leftarrow \phi - \gamma \nabla_\phi\sum_{i=1}^B \mathcal{L}_p(y_i, \alpha^i)$
    \Until{convergence} \\
    \Return $\phi$
    \end{algorithmic}
\end{algorithm}
\end{minipage}
\hfill
\begin{minipage}[t]{0.49\textwidth}
\begin{algorithm}[H]
    \caption{\\DHS: Data-driven Sampling}\label{algorithm2}
    Input: Dataset $\mathcal{D}$, $p$, $B$, $\gamma>0$, \textcolor{red}{$K \leq B$} \\
    Output: Model weights $\phi$
    
    \begin{algorithmic}[1]
    \Repeat
        \State Sample $\{y_1, y_2,...,y_B\} \sim \mathcal{D}$
            \State $\alpha \leftarrow \alpha^i \sim U[0,1]^p$ for $i=1,...,B$
            \State \textcolor{red}{$\tilde{\alpha} \leftarrow$ $\alpha$ sorted by $J(G_{H_\phi(\alpha)}, y_i)$}
            \State $\phi \leftarrow \phi - \gamma \nabla_\phi\sum_{i=1}^{\textcolor{red}{K}} \mathcal{L}_p(y_i, \textcolor{red}{\tilde{\alpha}^i})$
    \Until{convergence} \\
    \Return $\phi$
    \end{algorithmic}
\end{algorithm}
\end{minipage}
\caption{Algorithms for UHS and DHS training. Red text highlights differences between the two algorithms. In Line 4 of DHS, the sort is done in ascending order.}
\end{figure}

\subsubsection{Data-driven Hyperparameter Sampling (DHS):}
In the unsupervised setting, we propose to use the data-consistency loss induced by a setting of the regularization weights to assess whether the  reconstruction will be useful or not. Intuitively, values of $\boldsymbol{\alpha}$ which lead to high data-consistency loss $J(G_{H_\phi(\boldsymbol{\alpha})}, \boldsymbol{y})$ will produce reconstructions that deviate too much from the underlying anatomy, and which therefore can be ignored during training.

We propose the following strategy to exploit this idea. During training, one can compute gradients only on values of $\boldsymbol{\alpha}$ which induce a data-consistency loss below a pre-specified threshold. However, this presents two problems. First, calibrating and tuning this threshold can be difficult in practice. Second, at the beginning of training, since the main network will not produce good reconstructions, this threshold will likely not be satisfied. In lieu of a complex training/threshold scheduler, we instead adapt the threshold with the quality of reconstructions directly within the training loop.

Algorithm~\ref{algorithm2} outlines our proposed DHS strategy, which uses the best $K$ samples with the lowest data-consistency loss within a mini-batch of size $B$ to calculate the gradients. 
In effect, this dedicates model capacity for the subset of the hyperparameter landscape that can produce reconstructions that are most consistent with the data, while ignoring those that do not. The percentage of the landscape which is optimized is $K/B$.

\section{Experiments}

In all experiments, we restrict our attention to the case of $p=2$ regularization functions, although we emphasize that our method works for any number of regularization loss terms. 
We choose layer-wise total $\ell_1$-penalty on the weights~$\theta$ of the main reconstruction network and the total variation of the reconstruction image as two regularization loss terms:
\begin{align}
\begin{split}
    \mathcal{R}_1 &= \sum_{i=1}^L \norm{\hat{\theta}_i}_1, \\
\end{split}
\begin{split}
    \mathcal{R}_2 &= \norm{\nabla(G_{\hat{\theta}})}_1,
\end{split}
    \label{real-loss}
\end{align}
where $\hat{\theta} = H_\phi(\boldsymbol{\alpha})$, $\hat{\theta}_i$ denotes the weights of the reconstruction network for the $i$th layer, $L$ is the total number of layers of the reconstruction network, and $\nabla$ is the spatial gradient operator.

\subsubsection{Implementation}

The reconstruction network consists of a residual U-Net architecture \cite{unet} with 64 hidden channels per encoder layer, yielding a total of $n=592,002$ (non-trainable) parameters. The hypernetwork consists of fully-connected layers with leaky ReLU activations and batch normalizations between intermediate layers. Hypernetwork weights were initialized to approximate standard Kaiming initialization in the main network~\cite{Chang2020Principled}. We experiment with three hypernetwork architectures: ``small" consisting of 1-2-4-$n$ fully-connected layers, ``medium" consisting of 1-8-32-$n$ fully-connected layers, and ``large" consisting of 1-8-32-32-32-$n$ fully-connected layers.

For all models and baselines, we set the mini-batch size to $B=32$ and used ADAM optimizer with learning rate $1e-5$. For DHS, we took the top $25$\% of samples within a mini-batch, i.e. $K=8$, chosen using a grid-search, although we find that results are not sensitive to this optimal value. All training and testing experiments in this paper were performed on a machine equipped with an Intel Xeon Gold 6126 processor and an NVIDIA Titan Xp GPU, and all models were implemented in Pytorch.

\begin{table}[b]
\caption{Training and inference time for proposed models. Inference time is the runtime of one forward pass with a single $\boldsymbol{\alpha}$ and test measurement input, averaged over the test set. Mean $\pm$ standard deviation. ``$\sim$" denotes approximation.}
\begin{tabular*}{\textwidth}{@{\extracolsep{\fill}}ccccccccc}
\hline

 \textit{Hypernetwork} & \textit{Sampling Method} & \textit{Training time (hr)}  & \textit{Inference time (sec)} \\ \hline
 Baselines & - & $\sim$648 & 0.208$\pm$0.014 \\ 
\hline
 \multirow{2}{*}{Small} & UHS    & $\sim$6  &0.241$\pm$0.013 \\
& DHS & $\sim$7  & 0.237$\pm$0.016  \\
\hline
 \multirow{2}{*}{Medium} & UHS    & $\sim$12 & 0.256$\pm$0.023\\
& DHS & $\sim$15  & 0.259$\pm$0.021 \\
\hline
  \multirow{2}{*}{Large} & UHS    & $\sim$48 & 0.271$\pm$0.013\\
& DHS & $\sim$55 & 0.275$\pm$0.025\\
\hline
\end{tabular*}
\label{tab:runtime}
\end{table}

\begin{figure}[t]
\centering
\includegraphics[width=\textwidth]{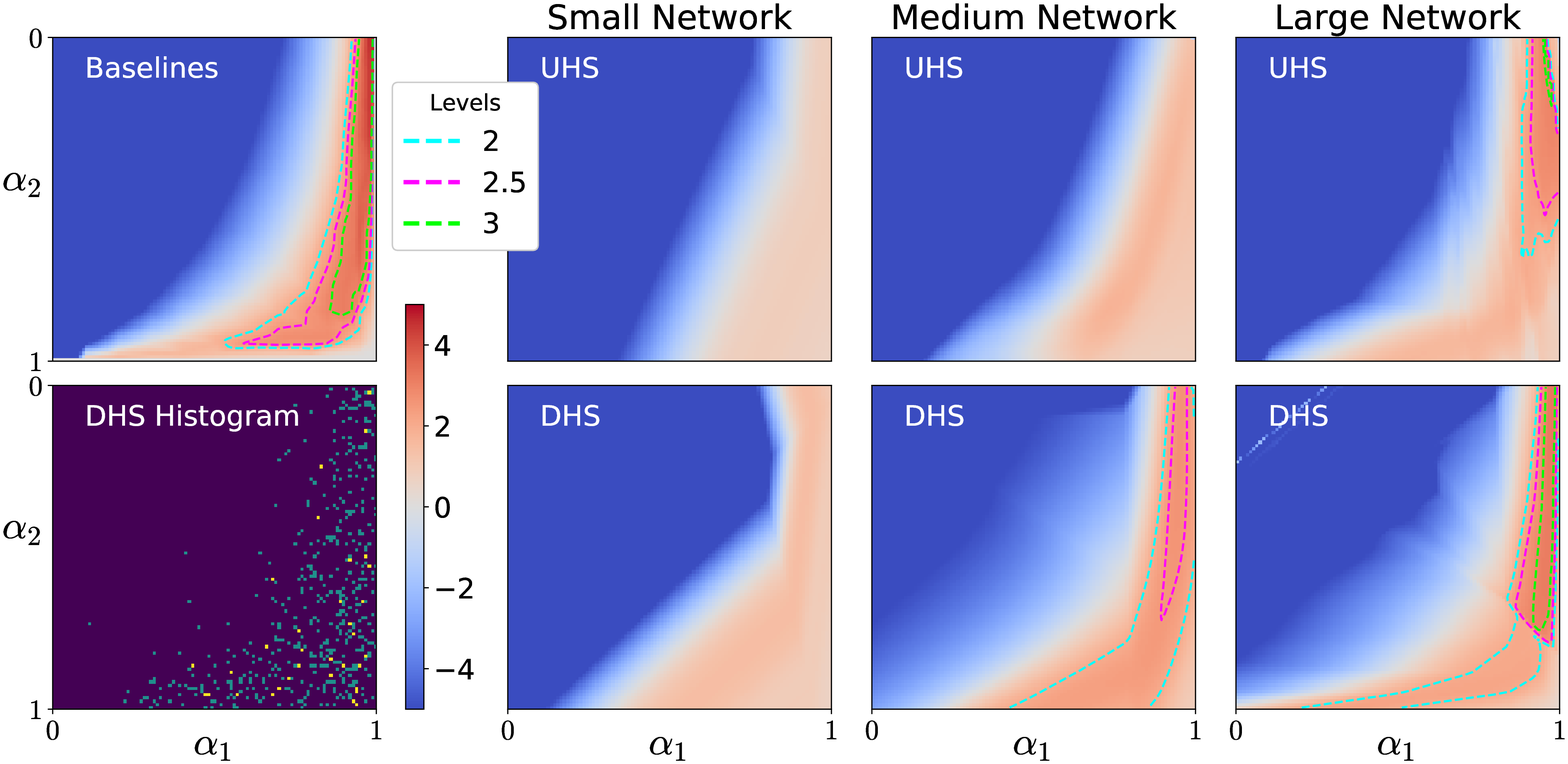}
\caption{RPSNR values over the hyperparameter support $[0,1]^2$ for different hypernetwork capacity and sampling methods. The $x$-axis and $y$-axis denote the value of the hyperparameters $\alpha_1$ and $\alpha_2$, respectively. Contours denote level sets of fixed value (see legend). (Left) The top image depicts the baseline landscape. The bottom image shows an example histogram of hyperparameter values used for gradient computation during one epoch of training with the DHS strategy. (Right) The top and bottom row show the UHS and DHS model landscapes, respectively, for three different hypernetwork capacities.} 
\label{fig:sampling}
\end{figure}

\subsubsection{Data.}

We trained and evaluated networks on a dataset of T1-weighted axial brain images~\cite{Dalca2018CVPR}. All images were intensity-normalized to the range $[0, 1]$ and cropped and re-sampled to a pixel grid of size $256 \times 256$. Dataset splits were $2000$, $500$, and $1000$ slices for training, validation, and testing. No subjects and scans were overlapping between the splits.
To generate the under-sampled measurements, we applied retrospective down-sampling with $4$-fold acceleration sub-sampling masks generated using a $2$nd-order polynomial Poisson-disk variable-density sampling strategy~\cite{Chauffert2013,Geethanath2013CompressedSM,lustig}.

\subsubsection{Evaluation Metrics.}

Reconstructions were evaluated against ground-truth full-resolution images on peak signal-to-noise ratio (PSNR). We constructed the \textit{relative} PSNR (RPSNR) value for a given reconstruction by subtracting the corresponding PSNR value for the zero-filled reconstruction, computed by applying the inverse Fourier transform directly to the k-space data where missing values are filled with zeros.

\subsubsection{Baseline Models.}
For comparison, we trained $324$ separate U-Net reconstruction networks for each fixed hyperparameter value, creating an $18\times 18$ grid within the hyperparameter space $[0,1]^2$. We refer to these models as benchmarks and emphasize that they demand significant computational cost, since each of these models need to be trained separately (see Table 1). We chose hyperparameter values non-uniformly to more densely sample in high RPSNR regions. 

\subsection{Hypernetwork Capacity and Hyperparameter Sampling}

We evaluate the performance of the proposed models using RPSNR landscapes over the space of permissible hyperparameter values $(\alpha_1, \alpha_2) \in [0,1]^2$. We generated landscapes for visualization by densely sampling the support $[0,1]^2$ to create a grid of size $100 \times 100$. For each grid point, we computed the value by passing the corresponding hyperparameter values to the model along with each under-sampled measurement $\boldsymbol{y}$ in the test set and taking the average RPSNR value. For baselines, the $18 \times 18$ grid was linearly interpolated to $100 \times 100$ to match the hypernetwork landscapes.

Fig.~\ref{fig:sampling} shows RPSNR landscapes for different reconstruction models. The top left map corresponds to baselines. The remaining images in the top row correspond to UHS models with varying hypernetwork capacities. The first image in the second row shows an example histogram of the regularization weight samples used for gradient computation over one DHS training epoch.  The rest of the second row shows RPSNR landscapes for DHS models at the same capacities. We observe that higher capacity hypernetworks approach the baseline models' performance, at the cost of computational resources and training time (see Table 1). We also observe significant improvement in performance using DHS as compared to UHS, given a fixed hypernetwork capacity. 

We find that the performance improvement achieved by DHS is less for the large hypernetwork, validating the expectation that the sampling distribution plays a more important role when the hypernetwork capacity is restricted.
In Fig.~\ref{fig:plot_rpsnr_threshold}a, we quantify these observations by plotting the percent area ($y$-axis) that exceed the corresponding RPSNR threshold ($x$-axis), for all models and baselines. Better-performing models have a larger area under the curve. 

\subsection{Range of Reconstructions}

\begin{figure}[t]
\centering
\includegraphics[width=\textwidth]{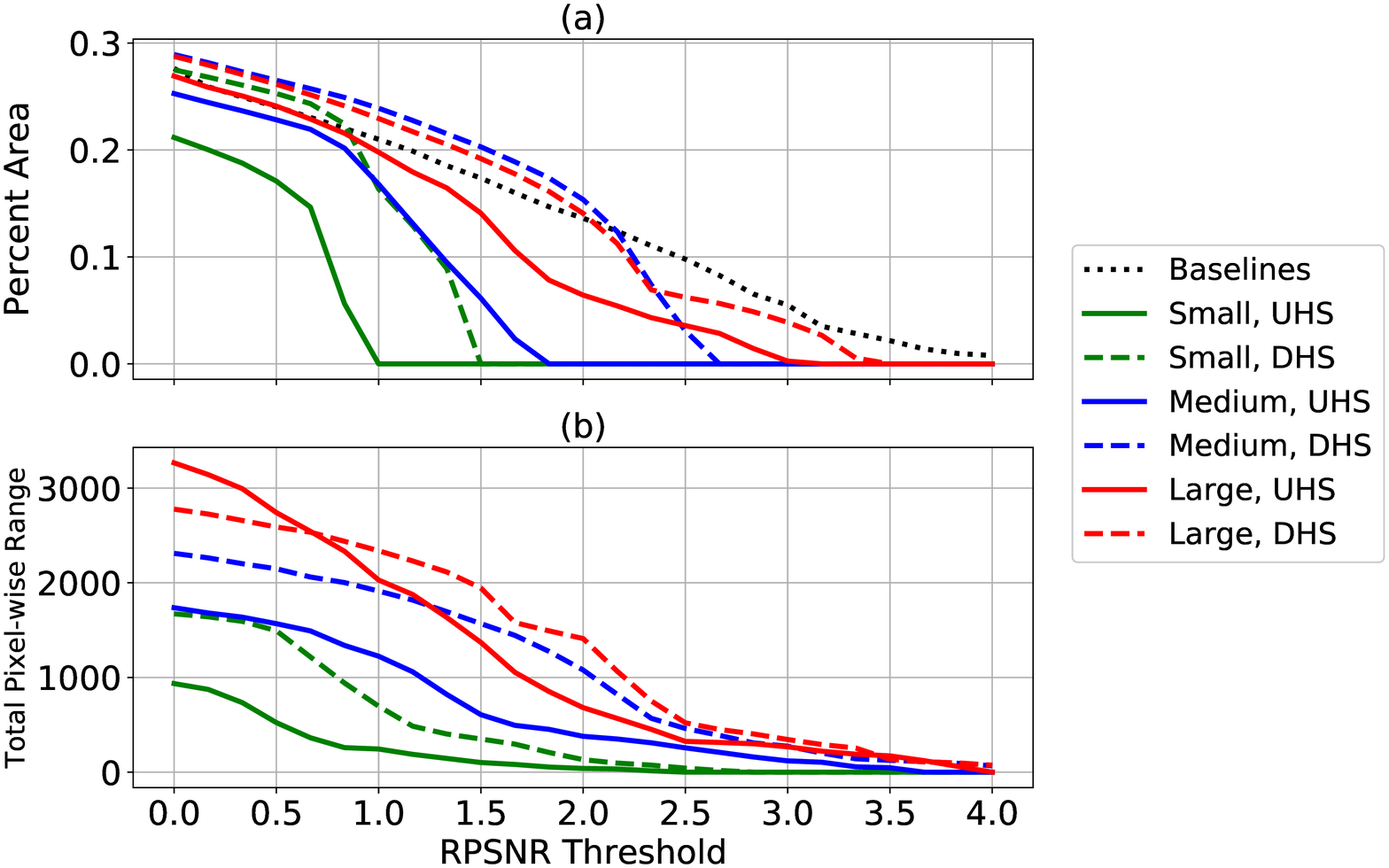}
\caption{(a) Percent of hyperparameter support area vs. RPSNR threshold. Values were calculated by computing the area of $[0,1]^2$ which exceeded the corresponding RPSNR value. (b) Total pixel-wise range vs. RPSNR threshold. Values were calculated by collecting all reconstructions exceeding the corresponding RPSNR value, computing the difference between the maximum and minimum pixel value for every pixel location across the set, and summing the result.} 
\label{fig:plot_rpsnr_threshold}
\end{figure}

Since our aforementioned workflow involved providing many different reconstructions for a user, we were interested in examining the range of reconstructions that were computed by the hypernetwork models.

Fig.~\ref{fig:plot_rpsnr_threshold}b shows the reconstruction range of different models.
We plot the total pixel-wise range ($y$-axis) in the set of reconstructions that exceed an RPSNR threshold ($x$-axis). A larger area under these curves is desirable, particularly for relatively high RPSNR thresholds.
We observe that higher capacity models and DHS yield a consistent increase in the range of reconstructions, compared to their lower capacity and UHS counterparts. 

\begin{figure}[!p]
\centering
\includegraphics[width=0.725\textwidth]{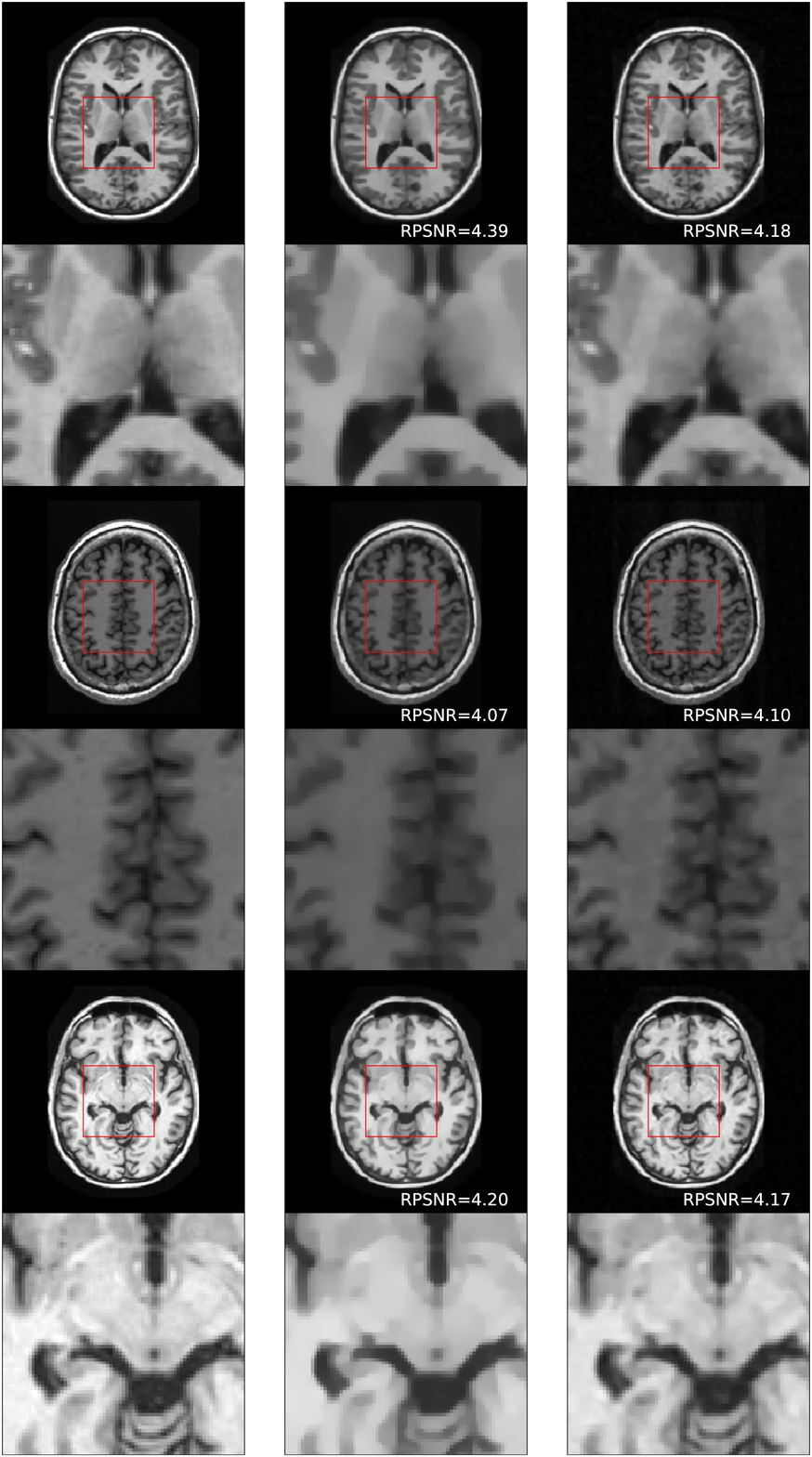}
\caption{Example slices, full-size and zoomed-in. First column are full-resolution images. Subsequent columns are reconstructions with similar RPSNR but computed with different regularization weights.} 
\label{fig:representative_slices}
\end{figure}

Fig. \ref{fig:representative_slices} shows some example reconstructions computed by the DHS large model. Along each row, we selected two reconstructions for the corresponding ground truth by collecting the subset of images within RPSNR range $[4, 4.5]$ and finding the two which had maximum distance in terms of $\ell_2$. In particular, we observe the varying levels of smoothness and sharp detail for a single slice despite the fact that the overall quality of the reconstructions (as measured by RPSNR) is virtually indistinguishable.

\section{Conclusion}
We introduce a novel unsupervised MR reconstruction method that is agnostic to regularization-weight hyperparameters in the amortized optimization formulation. We propose and demonstrate two hyperparameter sampling methods to maximize the performance of our proposed model, and show that it performs comparably to non-agnostic and computationally-expensive baselines. We furthermore highlight and quantify the range of reconstructions capable of being produced by our models. While our experiments focused on MRI reconstruction, this method can be applied broadly in order to render agnosticism towards any hyperparameter that may arise in many machine learning models. Furthermore, the reconstruction setting can be improved with more expressive networks and better regularization functions, particularly data-driven variants. 

\section{Acknowledgements}
This work was supported by NIH grants R01LM012719 (MS), R01AG053949 (MS), 1R01AG064027 (AD), the NSF NeuroNex grant 1707312 (MS), and the NSF CAREER 1748377 grant (MS).

\bibliographystyle{splncs04}
\bibliography{bib}

\end{document}